\documentclass[proceedings]{JHEP} 

\usepackage{epsfig}

\newbox\mybox
\newcommand\fverb{\setbox\mybox=\hbox\bgroup\verb}
\newcommand\fverbdo{\egroup\medskip\noindent\fbox{\unhbox\mybox}\ }
\newcommand\fverbit{\egroup\item[\fbox{\unhbox\mybox}]}

\font\beeg=cmr17 scaled 1600		
\newcommand\init[1]{\setbox\mybox=\hbox{{\beeg #1}~}%
		   \noindent\global\hangindent=\wd\mybox\global\hangafter-2%
		   \sc\smash{\llap {\lower 13.2pt \box\mybox}}}

\newcommand{\beq}{\begin{equation}}
\newcommand{\eeq}{\end{equation}}
\newcommand{\bea}{\begin{eqnarray}}
\newcommand{\eea}{\end{eqnarray}}
\newcommand{\ena}{\end{eqnarray}}
\renewcommand{\a}{\alpha}

\renewcommand{\d}{\delta}
\newcommand{\G}{\Gamma}
\newcommand{\e}{\epsilon}
\newcommand{\z}{\zeta}
\renewcommand{\l}{\lambda}

\newcommand{\Db}{\bar{D}}
\newcommand{\Phib}{\bar{\Phi}}

\newcommand{\SU}{\mathop{\rm SU}}
\newcommand{\SO}{\mathop{\rm SO}}

\conference{Quantum aspects of gauge theories, supersymmetry and unification,
Paris 1999}

\title{Correlation functions of chiral primary operators in perturbative
${\cal N}=4$ SYM}

\author{S. Penati $^a$, A. Santambrogio $^b$, D. Zanon $^c$\\
$^a$ Dipartimento di Fisica dell'Universit\`a di Milano--Bicocca, and\\ 
INFN, Sezione di Milano, Via Celoria 16,
20133 Milano, Italy\\
E-mail: \email{silvia.penati@mi.infn.it}
\\ \vspace{5pt} 
$^b$Instituut voor Theoretische Fysica - Katholieke Universiteit Leuven\\
Celestijnenlaan 200D B--3001 Leuven, Belgium\\
E-mail: \email{alberto.santambrogio@fys.kuleuven.ac.be}
\\ \vspace{5pt}
$^c$ Dipartimento di Fisica dell'Universit\`a di Milano, and\\ 
INFN, Sezione di Milano, Via Celoria 16,
20133 Milano, Italy\\
E-mail: \email{daniela.zanon@mi.infn.it}}

\abstract{We discuss recent results on two-point functions of chiral 
primary operators in ${\cal N}=4$ $\SU(N)$ supersymmetric Yang-Mills 
theory. Our results give further support to the belief that such correlators
are not renormalized to \emph{all}  orders in $g$ and to  \emph{all} 
orders in $N$.}
 
\begin{document}

{\init The aim} of this talk is to present recent results 
\cite{noi} on the explicit computation of two-point functions of chiral 
operators ${\rm Tr} \Phi^3$ in  ${\cal N}=4$ $\SU(N)$ supersymmetric  
Yang-Mills theory to
the  order   $g^4$  in  perturbation  theory, using  ${\cal N}=1$  
superspace techniques. We find that perturbative corrections to the 
correlators vanish for all $N$.  While
at order $g^2$ the cancellation \cite{HFS} can be ascribed to the
nonrenormalization theorem  valid for correlators of  operators in the
same multiplet as  the stress tensor, at order  $g^4$ this 
argument no longer  applies  and  the  actual  cancellation  occurs  
in a  highly nontrivial way, as will be shown.  

The plan of the talk is the following: after a brief introduction to 
the subject of correlation functions 
as tools to explore the AdS/CFT correspondence,
in section~\ref{s2} we briefly illustrate the ${\cal N}=4$ theory 
and give the relevant rules for calculating in ${\cal N}=1$ superspace.
In section~\ref{s3} we present the results of our calculations to order 
$g^4$: as a first test of our approach we check that the perturbative 
corrections to the two-point function with $k=2$ vanish, after that we 
consider the two-point correlator with $k=3$ and we describe all the 
order $g^4$ contributions.
We do not give technical details, for which we refer to our paper
\cite{noi}.

\section{Introduction}

Recently much evidence has been provided in testing the conjectured
equivalence of type $IIB$ superstring theory on anti-de-Sitter space
($ {\rm AdS}_5$) times a five--sphere to the ${\cal N}=4$ supersymmetric
$\SU(N)$ Yang-Mills conformal field theory living on the boundary, in
the large-$N$ limit and at large 't Hooft coupling $\lambda=g^2
N/4\pi$ ($g^2$ being the Yang-Mills coupling constant)~\cite{adscft}.
According to this correspondence correlation functions of operators in
the conformal field theory are mapped to appropriate on-shell
amplitudes of superstring theory in the bulk AdS background. 
\\
${\cal N}=4$ chiral primary operators 
\beq
{\rm Tr} \Phi^k\equiv {\rm Tr} \left( \Phi^{\{ i_1}(z)\Phi^{i_2}(z)
\cdots\Phi^{i_k\} }(z) \right) \,,
\eeq
in the symmetric, traceless representation of the R-symmetry group $\SU(4)$, 
play a special role in exploring non-perturbative 
statements concerning the above
mentioned connection. These are local operators of the lowest scaling
dimension in a given irreducible representation of the superconformal
algebra $\SU(2,2|4)$, and belong to short multiplets which are chiral
under a ${\cal N}=1$ subalgebra.  In the large-$N$ limit they
correspond to Kaluza Klein modes in the AdS supergravity sector.  In
the special case of $k=2$, two- and three-point correlators are given
by their free-field theory values for any finite $N$.  In this case
their form, fixed up to a constant by conformal invariance, is
protected by a nonrenormalization theorem~\cite{FGI} valid for two-
and three-point functions of operators in the same multiplet as the
stress tensor and as the $SU(4)$ flavor currents.

For any strong-weak coupling duality test it is essential to have
quantities that do not acquire radiative corrections as one moves from
weak to strong coupling.  If an exact computation in the supergravity
sector shows agreement with a tree level result in the Yang-Mills
sector, then there is an indication of a nonrenormalization theorem at
work.  This is the case for the three--point correlators $\langle{\rm Tr}
\Phi^{k_1}{\rm Tr} \Phi^{k_2}{\rm Tr} \Phi^{k_3}\rangle$ computed in
ref.~\cite{MS} in the large-$N$ limit of ${\cal N}=4$ $\SU(N)$
Yang-Mills: the strong limit result $\lambda=g^2 N/4\pi \gg 1$
obtained using type $IIB$ supergravity was shown to agree with the
weak 't Hooft coupling limit $\lambda=g^2 N/4\pi \ll 1$ in terms of
free fields. According to the AdS/CFT correspondence one concludes
that the correlators are independent of $\l$ to leading order in
$N$. A stronger conjecture made in ref.~\cite{MS} claims that
three-point functions might be independent of $g$ for \emph{any} value
of $N$.  As emphasized above, for the case $k=2$ nonrenormalization
properties have been proven to be enjoyed by two- and three-point
functions of chiral operators.  For general $k$ there exists evidence
of nonrenormalization based on proofs that rely on reasonable
assumptions (analyticity in harmonic superspace~\cite{EHW} and
validity of a generalized Adler-Bardeen theorem~\cite{PS}).

Explicit perturbative calculations in the ${\cal N}=4$ $\SU(N)$
Yang-Mills conformal field theory are a way to confirm the conjectures
and add insights into potential larger symmetries of the theory.
Important steps along this program have been made
in~\cite{HFS,S,HFMMR,BK}.  In particular, it has been shown that to
order $g^2$ radiative corrections do not affect the two- and
three-point functions of chiral operators~\cite{HFS}.  Two-point
functions have been computed for chiral operators with generic $k$
by showing that the order
$g^2$ contributions are proportional to the one for the $k=2$ case
which indeed
satisfies the known nonrenormalization theorem mentioned above. 
Concretely, the cancellation to order $g^2$ can be traced back to the fact
that at this order all the diagrams contain interactions involving at 
most two matter lines.
Clearly this is not true, for example, at order $g^4$, where diagrams
with gluon exchanges among three matter lines appear.
Therefore, it
is interesting to investigate whether the cancellation shown in \cite{HFS} 
for the $k > 2$ case is an accident of order $g^2$.

In our paper \cite{noi} we have addressed the nontrivial test left 
open at order $g^4$, by computing the two-point function for the operator
${\rm Tr} \Phi^k$ in the case $k=3$. The analysis for generic $k$ 
is now under investigation \cite{noinew}. However, as already mentioned,
at order $g^4$ the $k=3$ case is a crucial test, being diagrams with 
interactions involving three matter lines present.
We have found that corrections indeed vanish for \emph{all} values of $N$, 
then supporting the stronger conjecture of ref.~\cite{MS}.

\section{The main features of our calculation}\label{s2}

The physical particle content of ${\cal N}=4$ supersymmetric
Yang-Mills theory is given by one spin-$1$ vector, four
spin-${1}/{2}$ Majorana spinors and six spin-$0$ particles in the
${\bf 6}$ of the R-symmetry group $\SU(4)\sim\SO(6)$. All particles are 
massless and transform under the adjoint representation of the $\SU(N)$ gauge
group.

Perturbative calculations are quite difficult to handle using a
component field formulation of the theory. (Note that in
ref.~\cite{HFS} a component approach was used, but the order $g^2$
result for the two- and three-point correlators was obtained using a
general argumentation based on colour combinatorics.  Only a schematic
knowledge of the structure of the component action was required.) In
general, in order to resum Feynman diagrams at higher-loop orders it
is greatly advantageous to work in superspace.

In ${\cal N}=1$ superspace the action can be written in terms of one
vector superfield $V$ (real) and three chiral superfields $\Phi^i$
containing the six scalars organized into the ${\bf 3}\times \bf{ \bar
3}$ of $\SU(3) \subset \SU(4)$ (we follow the notations
in~\cite{superspace})
\bea
S[J,\bar{J}]
&=&\int d^8z~ {\rm Tr}\left(e^{-gV} \Phib_i e^{gV} \Phi^i\right)
\nonumber\\
&&+\frac{1}{2g^2}\int d^6z~ {\rm Tr} W^\a W_\a
\nonumber\\
&&+\frac{ig}{3!} {\rm Tr} \int d^6z~ \e_{ijk} \Phi^i
[\Phi^j,\Phi^k]
\nonumber\\
&&+\frac{ig}{3!} {\rm Tr} \int d^6\bar{z}~ \e_{ijk} \Phib^i
[\Phib^j,\Phib^k]
\nonumber\\
&&+\int d^6z~ J {\cal O}+\int d^6\bar{z}~ \bar{J}\bar{{\cal O}}\,,
\label{actionYM}
\eea
where $W_\a= i\Db^2(e^{-gV}D_\a e^{gV})$, and $V=V^aT^a$,
$\Phi_i=\Phi_i^a T^a$, $T^a$ being $N\times N$ matrices in the
fundamental representation of $\SU(N)$.  We have added to the classical action
source terms for the chiral primary operators generically denoted by
${\cal O}$ since our goal is the computation of their correlators.

Although in~(\ref{actionYM}) the ${\cal N}=4$ supersymmetry invariance
is realized only non linearly, the main advantage offered by a ${\cal
N}=1$ formulation of the theory resides in the fact that a
straightforward off-shell quantum formulation is available. Thus if
the aim is to perform higher-loop perturbative calculations this is
the most suited approach to follow. Feynman rules are by now standard
(we refer to appendix B of \cite{noi} for a complete list).

We will now focus on the two-point super-correlator for the operator
${\cal O}={\rm Tr}(\Phi^{\{ i} \Phi^j \Phi^{k\} })$.  As in
ref.~\cite{HFS}, we consider the $\SU(3)$ highest weight $\Phi^1$
field and compute $\langle{\rm Tr}(\Phi^1)^3{\rm
Tr}(\Phib^1)^3\rangle$. This is not a restrictive choice since all the
other primary chiral correlators can be obtained from this one by
$\SU(3)$ transformations.  What we gain is that we have no flavour
combinatorics and we are left to deal with the colour combinatorics
only.

We work in euclidean space, with the generating functional defined as
\beq
W[J,\bar{J}]=\int {\cal D}\Phi~{\cal D}\Phib\,{\cal D}V\,e^{S[J,\bar{J}]}\,.
\label{genfunc}
\eeq
Thus the two-point function is given by
\beq
\langle {\rm Tr}(\Phi^1)^3(z_1){\rm Tr}(\Phib^1)^3(z_2)\rangle =
\left. \frac{\d^2 W}{\d J(z_1)\d\bar{J}(z_2)}\right|_{J,\bar{J}=0}
\label{defcorr}
\eeq
where $z \equiv (x,\theta, \bar{\theta})$.  We use perturbation theory
to evaluate the contributions to $W[J,\bar{J}]$ which are quadratic in
the sources, i.e. of the form
\beq
\int d^4x_1~d^4x_2~ d^4\theta\, 
J(x_1,\theta,\bar{\theta}) \frac{F(g^2,N)}{(x_1-x_2)^6}
\bar{J}(x_2,\theta,\bar{\theta})\,,
\label{twopoint}
\eeq
where the $x$-dependence of the result is fixed by the conformal
invariance of the theory, and the function $F(g^2,N)$ is what we want
to determine up to order $g^4$. We will find a result valid for any
$N$.

In order to perform the calculation we have found it convenient to
work in momentum space, using dimensional regularization and minimal
subtraction scheme.  In $n$ dimensions, with $n=4-2\e$, the Fourier
transform of the power factor $(x_1-x_2)^{-6}$ in~(\ref{twopoint}) is
given by
\beq
\frac{1}{(x^2)^3}=
\frac{\pi^{-2+\e}}{64} \frac{\G(-1-\e)}{\G(3)}
\int d^n p ~\frac{e^{-ipx}}{(p^2)^{-1-\e}}\,.
\label{basicformula}
\eeq
The presence of the singular factor $\G(-1-\e) \sim {1}/{\e}$ signals,
in momentum space and in dimensional regularization, the UV divergence
of the correlation function in~(\ref{twopoint}) associated to the
short-distance behaviour for $x_1 \sim x_2$.  It follows that
performing perturbative calculations in momentum space it is
sufficient to look for all the contributions to~(\ref{twopoint}) that
behave like $1/\e$, therefore disregarding finite contributions.  In
fact, once the divergent terms are determined at a given order in $g$,
using (\ref{basicformula}) one can reconstruct an $x$-space structure
as in (\ref{twopoint}) with a non-vanishing contribution to
$F(g^2,N)$.  Finite contributions in momentum space would correspond
in $x$-space to terms proportional to $\e$ which give rise only to
contact terms~\cite{CT}.

The one stated above is the basic rule of our strategy that
we can summarize as follows:
\begin{itemize}
\item consider all the two-point diagrams from $W[J,\bar{J}]$ with $J$
and $\bar{J}$ on the external legs,
\item evaluate all the factors coming from combinatorics of the diagram
and compute the colour structure,
\item perform the superspace $D$-algebra following standard
techniques,
\item reduce the result to a multi-loop momentum integral,
\item compute its $1/\e$ divergent contribution.
\end{itemize}
This last step, i.e.\ the calculation of the divergent part of the
various integrals we have achieved using the method of
uniqueness~\cite{kazakov} and various rules and
identities~\cite{CT,russians} that we have collected in
appendix B of \cite{noi}. Since the theory is at its conformal point, it is
not affected by IR divergences. Therefore, even if we work in a
massless regularization scheme, we never worry about the IR behavior
of our integrals. Moreover, since the theory is finite, the diagrams
that we consider do not possess UV divergent subdiagrams.  Finally, as
a general remark we observe that gauge-fixing the classical action
requires the introduction of corresponding Yang-Mills ghosts. However
they only couple to the vector multiplet and do not enter our specific
calculation.

In the next section we will apply the general procedure just described
to the analysis of the two-point function $\left<{\rm Tr}(\Phi^1)^k{\rm Tr}
(\Phib^1)^k\right>$ with $k=3$ to order $g^4$.

\section{Correlation functions to order $g^4$}\label{s3}

Before coming to our main calculation, the $k=3$ case, we will first sketch
how our formalism works in a simpler case, the order $g^4$ calculation of the
two-point correlator with $k=2$. As previously discussed, in this case we
already know that perturbative corrections should not be there: this simpler
calculation is then a non trivial test of our techniques.

The two-point correlator we are interested in is obtained from
$W[J,\bar{J}]$ inserting in the action~(\ref{actionYM}) the chiral
operators ${\cal O}={\rm Tr} (\Phi^1)^2$ and $\bar{{\cal O}}={\rm
Tr}(\Phib^1)^2$. As outlined in the previous section, the relevant
contribution is obtained from the generating functional isolating
terms of the form
\beq
\int d^4x_1\,d^4x_2\, d^4\theta\, J(x_1,\theta,
\bar{\theta})\frac{E(g^2,N)}{(x_1-x_2)^4}\bar{J}(x_2,\theta,\bar{\theta})\,.
\label{twopointtwo}
\eeq

\EPSFIGURE{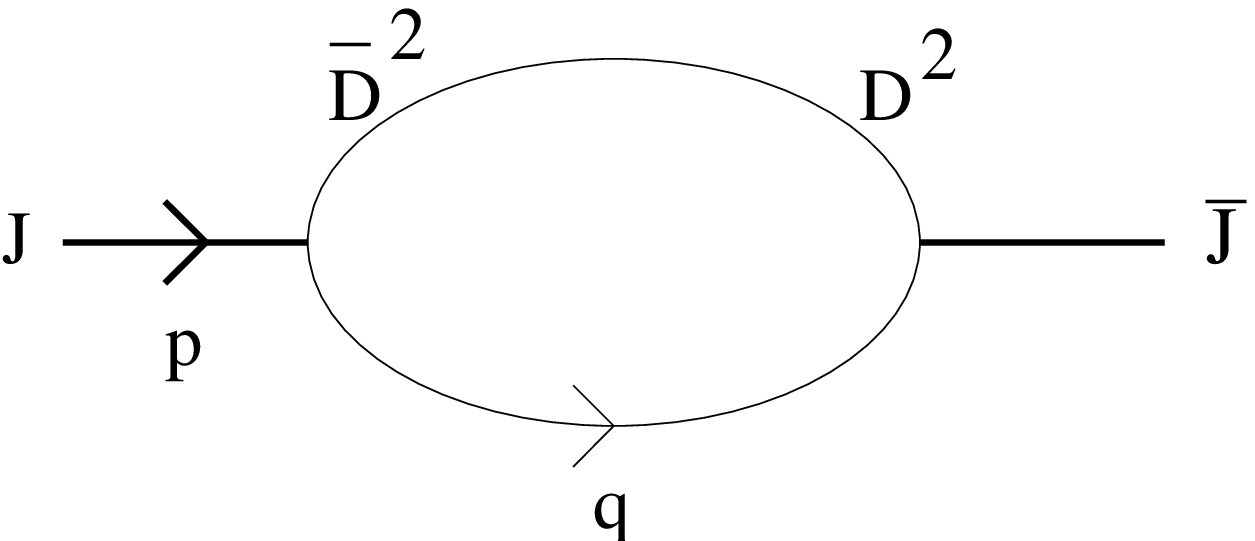, width=15em}{Tree-level contribution to $\langle {\rm
Tr}(\Phi^1)^2 {\rm Tr}(\Phib^1)^2 \rangle $.\label{f1}}

\noindent The general form of~(\ref{twopointtwo}) is fixed by conformal invariance, while
the function $E(g^2,N)$ is the unknown to be determined. Fourier transforming
from $x$-space to momentum space
\beq
\frac{1}{(x^2)^2}=\frac{\pi^{-2+\e}}{16} \frac{\G(-\e)}{\G(2)}
\int d^n p ~\frac{e^{-ipx}}{(p^2)^{-\e}}
\label{basicformulatwo}
\eeq
makes it clear that non-trivial contributions to the generating
functional are given by the divergent part of our Feynman diagrams.

To start with we consider the tree-level contribution corresponding to
the graph in figure~\ref{f1}. The calculation in this case is very simple
\cite{noi}: its contribution to the two-point function is given by 
\beq
\frac{1}{(4\pi)^2}\,2(N^2-1)\,\frac{1}{\e}
\int d^4p\,d^4\theta \, J(-p,\theta,\bar{\theta})\bar{J}(p,\theta,\bar{\theta})\,.
\label{treetwo}
\eeq

\EPSFIGURE{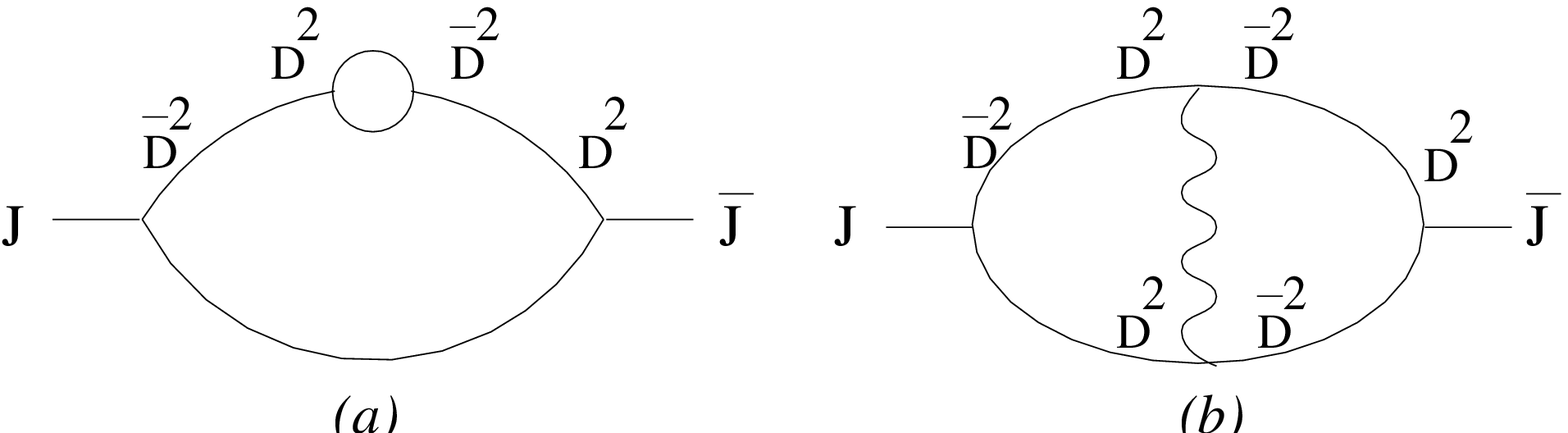, width=24em}{$g^2$-order contribution to $\langle {\rm
Tr}(\Phi^1)^2 {\rm Tr}(\Phib^1)^2\rangle $.\label{f2}}

The order $g^2$ contribution, once evaluated in superspace gives
immediately a zero result: the diagrams one would need to consider are
shown in figure~\ref{f2}.  Diagram~\ref{f2}$a$ does not contribute since the one-loop
correction to the chiral propagator vanishes due to a complete
cancellation between vector and chiral loops~\cite{GSR}. Diagram~\ref{f2}$b$,
after completion of the $D$-algebra leads to a finite momentum
integral.

Now we consider the order $g^4$ contributions: they are shown in 
figure~\ref{f3}.

\EPSFIGURE[t]{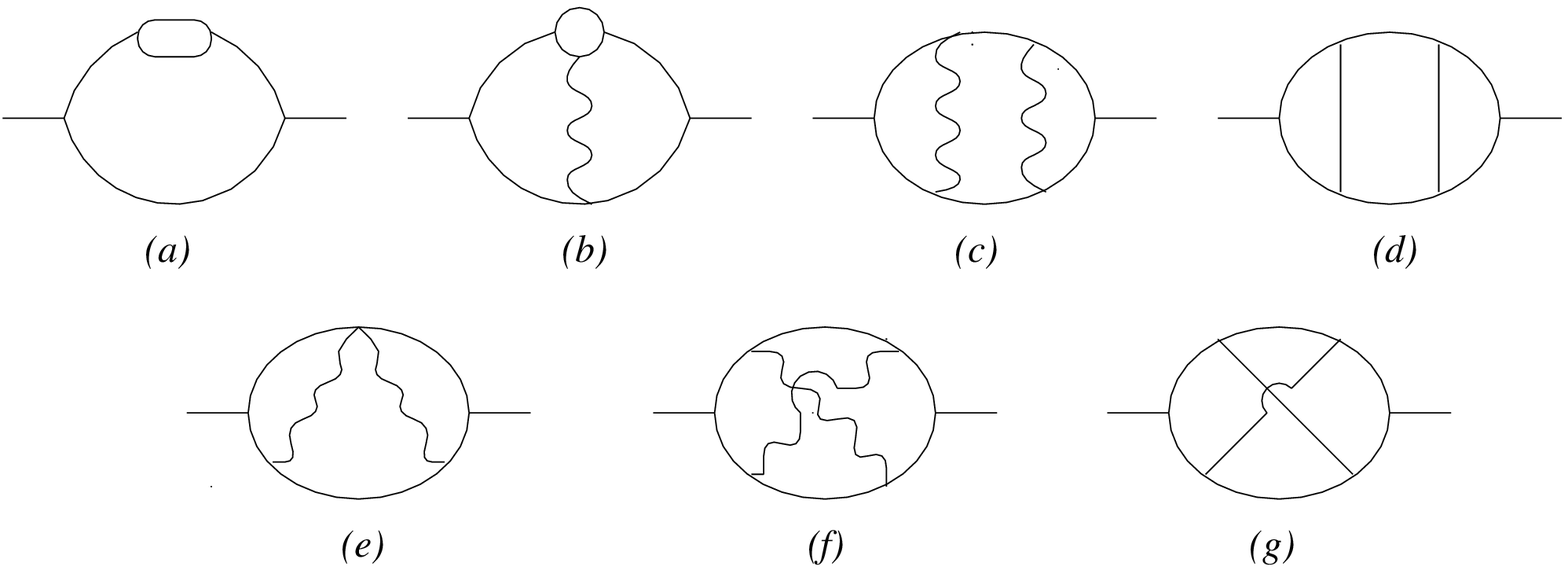,width=36em}{$g^4$-order contribution to 
$\langle {\rm Tr}(\Phi^1)^2 {\rm Tr} (\Phib^1)^2\rangle $.\label{f3}}

In figure~\ref{f3}$a$ we have the insertion of a two-loop propagator
correction, while in figure~\ref{f3}$b$ a one-loop vertex correction 
appears~\cite{GSR}.
Note that a diagram with a vector propagator corrected at order $g^2$
is absent since at one-loop order there is a complete cancellation
among chiral, vector and ghost contributions~\cite{GSR}.

The graph in figure~\ref{f3}$a$ is easy to compute: with an overall factor
\bea
&&~~~~~~~~~\frac{16}{(4\pi)^6}g^4\,N^2(N^2-1)\nonumber\\
&&\times\int d^4p\,d^4\theta\,J (-p,\theta,\bar{\theta}) 
\bar{J}(p,\theta,\bar{\theta})
\label{overall}
\eea
one obtains the following divergent contribution
\beq
{\rm figure~\ref{f3}}a \rightarrow \zeta(3)\frac{1}{\e}\,.
\label{3a}
\eeq
For figure~\ref{f3}$b$, with the same overall factor as in~(\ref{overall}),
one obtains
\beq
{\rm figure~\ref{f3}}b \rightarrow -\,2\zeta(3)\frac{1}{\e}\,.
\label{3b}
\eeq

A rather straightforward computation of the $D$-algebra for the diagrams in
figures~\ref{f3}$c$,~\ref{f3}$d$ and \ref{f3}$e$ allows to
conclude that the corresponding momentum integrals are
actually all finite and, as previously observed, not relevant for our purpose.

Finally, for the graphs in figure~\ref{f3}$f$ and in figure~\ref{f3}$g$,
factoring out the same overall quantity we have
\beq
{\rm figure~\ref{f3}}f \,\rightarrow \,\frac{1}{2}\zeta(3)\frac{1}{\e}\,,
\label{3f}
\eeq
and
\beq
{\rm figure~\ref{f3}}g \,\rightarrow \,\frac{1}{2}\zeta(3)\frac{1}{\e}\,.
\label{3g}
\eeq

It is a trivial matter to sum up the contributions listed
in~(\ref{3a}), (\ref{3b}), (\ref{3f}) and (\ref{3g}) and obtain a
vanishing result, as expected from the nonrenormalization theorem.
\pagebreak[3]

We note that the diagrams in figures \ref{f3}$f$
and \ref{f3}$g$ lead to planar contributions, i.e. with exactly the same $N$
dependence from colour combinatorics as the other diagrams (the $N$ dependence
is the one shown in the common overall factor (\ref{overall})): indeed to
this order nonplanar diagrams are absent.  
In the $k=3$ case we will be confronted with a more complicated situation.

\vskip 5mm

Let us now come to the computation of the two-point function for the
chiral operator ${\cal O}={\rm Tr}(\Phi^1)^3$. To this end we go back
to~(\ref{twopoint}) and compute the perturbative contributions to the
function $F(g^2,N)$.  As previously emphasized, making use
of~(\ref{basicformula}) we write Feynman diagrams in momentum space
and isolate the $1/\e$ poles.

\EPSFIGURE[b]{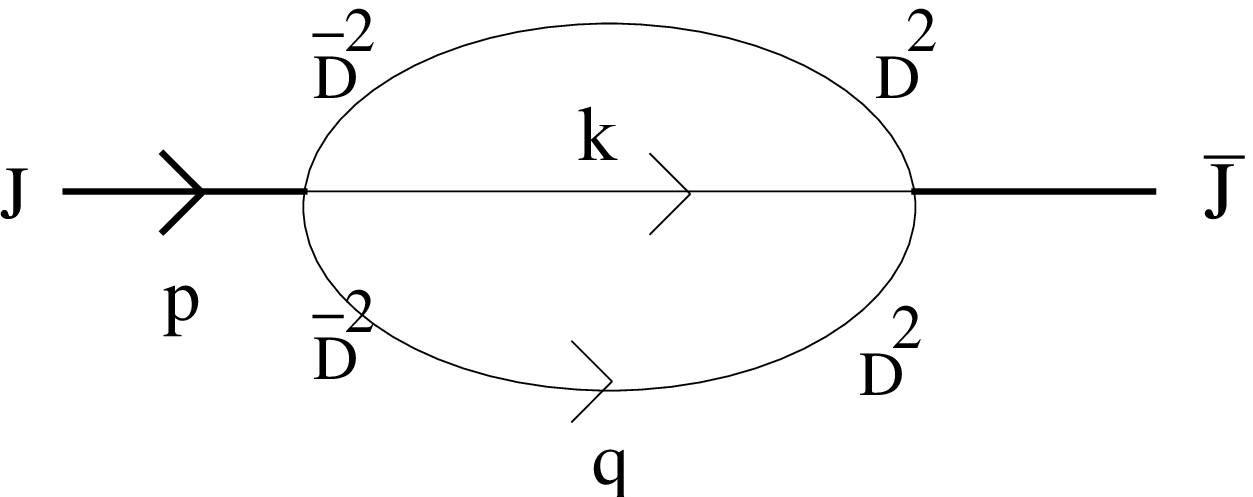, width=16.9em}{Tree-level contribution to $\langle
{\rm Tr}(\Phi^1)^3 {\rm Tr}(\Phib^1)^3\rangle $.\label{f4}}

In figure~\ref{f4} we have drawn the tree-level contribution. 
With an overall factor
\bea
&&~~~~~\frac{3}{(4\pi)^4}\,\frac{(N^2-1)(N^2-4)}{N}\nonumber\\
&&\times\int d^4p\,d^4\theta\,J(-p,\theta,\bar{\theta})
\bar{J}(p,\theta,\bar{\theta})
\eea
we obtain
\beq
{\rm figure~\ref{f4}} \rightarrow -\frac{1}{4\e}p^2\,.
\label{3tree}
\eeq
The result in $x$-space is readily recovered using formula 
(\ref{basicformula}).

\EPSFIGURE[t]{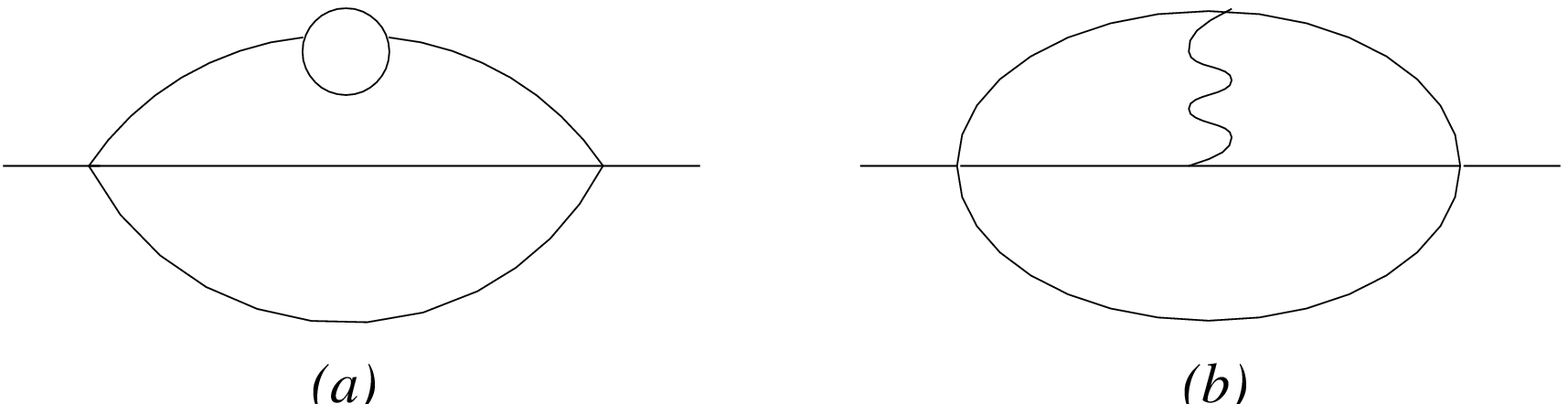, width=24em}{$g^2$-order  contribution to $\langle {\rm
Tr}(\Phi^1)^3 {\rm Tr} (\Phib^1)^3\rangle $.\label{f5}}

The superspace diagrams that enter the order $g^2$ computation are
shown in figure~\ref{f5}.  They are nothing but the ones that
appear in figure~\ref{f2} with one line added from the chiral external
vertices. One proves that their contributions vanish with exactly the
same reasoning outlined previously. As found in
ref.~\cite{HFS} to order $g^2$ the vanishing of the correlator is due
to the fact that it is proportional to the correlator of ${\cal
O}={\rm Tr}(\Phi^1)^2$ for which the nonrenormalization theorem is
valid.  However, this is no longer true at order $g^4$ to which we
turn now.

\EPSFIGURE[t]{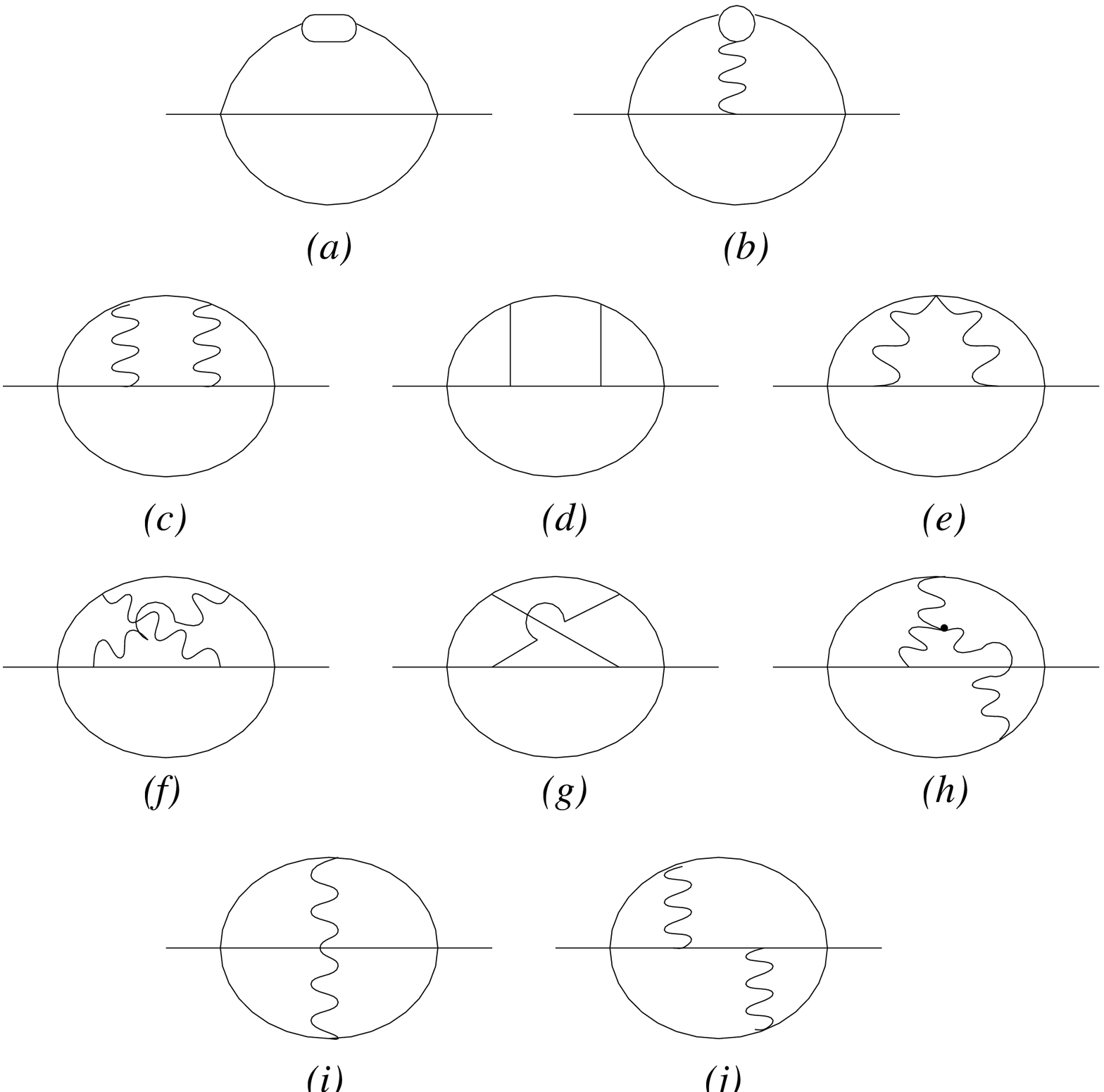, width=24.5em}{$g^4$-order contribution to
$\langle {\rm Tr}(\Phi^1)^3 {\rm Tr}(\Phib^1)^3\rangle $.\label{f6}}

The diagrams contributing to $g^4$-order are collected in
figure~\ref{f6}.  The ones in figure~\ref{f6}$a$--\ref{f6}$g$ are the
same as in figure~\ref{f3} with one extra line added from the chiral
external vertices. From the result obtained in the previous case at
order $g^4$, we would be tempted to believe
that these diagrams still sum up to zero. However this would be a
wrong conclusion.  In fact, what makes things different is that
in~\ref{f6}$f$ and~\ref{f6}$g$ the addition of the extra line changes
completely the topology of the diagrams which become really
\emph{nonplanar}. As a consequence, their colour combinatorics changes
and their $N$-dependence is distinct from the remaining planar
diagrams~\ref{f6}$a$--~\ref{f6}$e$. More specifically, in this case it
turns out that the nonplanar diagrams~\ref{f6}$f$ and~\ref{f6}$g$ lead
to a vanishing colour combinatorics factor.

The evaluation of the colour coefficient for the other nonplanar
diagram in figure~\ref{f6}$h$ reveals again a vanishing contribution.
The fact that the nonplanar
diagrams do not contribute indicates that the final answer is going to
be valid for all values of $N$, independently of any large-$N$
limit. In light of this result it becomes challenging to prove the
cancellation of nonplanar diagrams to all orders in the Yang-Mills
coupling. Moreover it is natural to ask if this mechanism of cancellation
is still valid for two-point correlation functions of the form 
$\langle{\rm Tr}(\Phi^1)^k{\rm Tr}(\Phib^1)^k\rangle$, with $k>3$. However,
a simple direct analysis shows that this is not true \cite{noinew}. 

Going back to figure~\ref{f6}, one easily convinces oneself that for
the graphs in figures~\ref{f6}$c$,~\ref{f6}$d$ and~\ref{f6}$e$ the
same analysis as in the previous section applies.  In this case the
addition of the chiral line simply adds a $D^2\Db^2$ factor which
accounts for the $D$-algebra of one added loop; performing the
$D$-algebra in the diagrams one is left with finite integrals.

We note that at this order
diagrams containing the scalar superpotential vertex
\beq
\e_{ijk} {\rm Tr} (\Phi^{i} [\Phi^{j},\Phi^{k}]) 
\eeq
do not contribute.

We are left with the contributions from
figures~\ref{f6}$a$,~\ref{f6}$b$,~\ref{f6}$i$ and~\ref{f6}$j$. We will 
find that a highly nontrivial cancellation occurs.

For every diagram we need compute the specific combinatorics, the
various factors from vertices and propagators and the colour structure. Then
we have to perform the $D$-algebra in the loops  and finally evaluate the
momentum integrals. We factorize for each contribution the same quantity
\bea
&&~~\frac{9}{(4\pi)^8}\,g^4 ~N(N^2-4)(N^2-1)\nonumber\\
&&\times\int d^4p\,d^4\theta\,
J(-p,\theta,\bar{\theta}) \bar{J}(p,\theta,\bar{\theta})\,.
\label{relevantoverall}
\eea
The diagram in figure~\ref{f6}$a$ which contains the two-loop propagator 
correction, gives
\beq
{\rm figure~\ref{f6}}a
\rightarrow  -\frac{3}{2}\z(3)\frac{1}{\e}\,p^2\,.
\label{6a}
\eeq
The diagram in figure~\ref{f6}$b$ contains the one-loop vertex
correction. In this case the resulting contribution is given by
\beq
{\rm figure~\ref{f6}}b\,\rightarrow\,3\z(3)\frac{1}{\e}\, p^2\,.
\label{6b}
\eeq
In the same way for the graph in figure~\ref{f6}$i$ one has
\beq
{\rm figure~\ref{f6}}i \rightarrow\,-5\z(5)\frac{1}{\e}\,p^2\,.
\label{6i}
\eeq
Finally we concentrate on the diagram in Fig.~$6j$. The evaluation 
of the corresponding
momentum integral is highly nontrivial  and we refer to our paper \cite{noi} 
for all the technical details. The result is given by
\beq
{\rm figure~\ref{f6}}j \,\rightarrow
\left[5\z(5)-\frac{3}{2}\z(3)\right]\frac{1}{\e}\,p^2\,.
\label{6j}
\eeq
At this point it is simple to add the four contributions in
(\ref{6a}),~(\ref{6b}),~(\ref{6i}) and~(\ref{6j}) and check the
complete cancellation of the $1/\e$ terms. It is interesting to note
that, while the diagrams $6a, ~6b$ only contribute with a divergent
term proportional to $\z(3)$ and the diagram $6i$ gives only a
$\z(5)$-term, from the diagram $6j$ both terms arise with the correct
coefficients to cancel completely the divergence.

\section{Conclusions}\label{s4}

We have discussed the calculation of the two-point correlation function 
for the chiral primary operator ${\rm Tr}\Phi_1^3$ in ${\cal N} =4$ 
$\SU(N)$ SYM theory up to $g^4$-order. We have found a
complete cancellation of quantum corrections for any finite $N$.  Our
result represents the first ${\cal O}(g^4)$ direct check of the
nonrenormalization theorem conjectured on the basis of the AdS/CFT
correspondence~\cite{MS}. It supports also the stronger
claim~\cite{MS} that there might be no quantum corrections at all, for
any finite $N$.

We have performed the calculation in ${\cal N}=1$ superspace using
dimensional regularization. The loop-integrals have been evaluated in
momentum space with the method of uniqueness~\cite{kazakov, russians}.
In momentum space nontrivial, potential contributions appear as local
divergent terms that are easily isolated and evaluated. Finite
contributions would correspond to contact terms and can be neglected.

Our procedure is applicable to the perturbative analysis of more
complicated cases. Two-point functions for ${\rm Tr} \Phi^k$, $k>3$,
three-point functions and extremal correlators for chiral primary
operators are now under investigation \cite{noinew}.

\acknowledgments

This work has been supported by the European Commission TMR programme
ERBFMRX-CT96-0045, in which S.P. and D.Z. are associated to the University of 
Torino.

\end{document}